# Low-temperature, *in situ* tunable, uniaxial stress measurements in semiconductors using a piezoelectric actuator

M. Shayegan,[1,2] K. Karrai,[1] Y. P. Shkolnikov,[2] K. Vakili,[2] E. P. De Poortere,[2] and S. Manus[1]

[1] Center for Nanoscience, Sektion Physik der Ludwig-Maximilians-Universitaet, Geschwister-Scholl-Platz 1, 80539 Muenchen, Germany

[2] Department of Electrical Engineering, Princeton University, Princeton, NJ 08544

**Abstract**

We demonstrate the use of a piezoelectric actuator to apply, at low-temperatures, uniaxial stress in the plane of a two-dimensional electron system confined to a modulation-doped AlAs quantum well. Via the application of stress, which can be tuned *in-situ* and continuously, we control the energies and occupations of the conduction-band minima and the electronic properties of the electron system. We also report measurements of the longitudinal and transverse strain versus bias for the actuator at 300, 77, and 4.2K. A pronounced hysteresis is observed at 300 and 77K, while at 4.2K strain is nearly linear and shows very little hysteresis with the applied bias.

Electronic properties of semiconductors can be strongly modified via the application of stress. Using uniaxial stress, e.g., one can induce a splitting of the conduction-band energy minima (valleys) in semiconductors with multi-valley occupancy [1]. Since such measurements are traditionally performed using a vice, the *in-situ* tuning of stress at low temperatures has proved to be challenging. Here we describe *in-situ*-tunable uniaxial stress measurements at cryogenic temperatures, using a piezoelectric (piezo) actuator, on a high-mobility two-dimensional (2D) electron system confined to a modulation-doped AlAs quantum well. Using the uniaxial stress, we continuously change the energies and occupations of different conduction-band valleys and therefore the electronic properties of the 2D system. Given the simplicity and versatility of our technique, it should find widespread use in studies of uniaxial stress measurements in thin-film structures, including low-dimensional semiconductor systems. We also provide a detailed characterization of the longitudinal and transverse strain of this piezo actuator, as a function of bias, in a large temperature range. This characterization is useful not only for the measurements we report, but also for the design and operation of widely used instruments, such as scanning probe microscopes, at low temperatures [2].

Before presenting uniaxial stress measurements in AlAs 2D electrons, we first describe our characterization of the actuator we used, and demonstrate that, at low temperatures (77K and below), the piezo strain can be fully transmitted to the sample. In our experiments, we used a stacked PZT ($PbZrTiO_3$) piezo rod [3] with an active length of 7mm and a 5x5mm$^2$ cross-section. To characterize the strain, we glued two resistance strain gauges [4], using a two-component epoxy [5], to the opposite, flat faces of the piezo rod [see Fig. 1(c) inset]. One of the gauges was mounted so that it measured the strain along the piezo rod's poling direction (*x* direction), while the other measured strain perpendicular to this direction (*y* direction). The piezo was then placed in a cryostat, evacuated, and partially filled with He exchange gas. For 77 and 4.2K measurements the cryostat was immersed in liquid N and He, respectively. We used a homemade AC resistance bridge circuit, with a low input-resistance current amplifier, to measure the strain gauge resistance. To achieve the stability required for precision measurements, we found it best to measure the gauge resistance against the resistance of another, free-standing gauge which was placed nearby in the cryostat and cooled to the same temperature. We were able to measure strain with a typical absolute accuracy of 5%.

In Fig. 1 we present our results at three different temperatures [6]. Plotted is the strain, defined as $\Delta L/L$, the percentage change in the size $L$ of the piezo in the indicated direction. Data are shown for both increasing and decreasing bias across the piezo, and also for the size changes induced in the $x$ and $y$ directions [$(\Delta L/L)_x$ and $(\Delta L/L)_y$]. Note that, as expected, $(\Delta L/L)_y$ has the opposite sign of $(\Delta L/L)_x$. The data were taken by incrementing (or decrementing) the piezo bias by 10V steps, waiting for 2 to 3 minutes until $\Delta L/L$ changed by less than about 2% with time (negligible creep), and then recording the measured $\Delta L/L$.

Two main features of Fig. 1 data are noteworthy. First, there is significant hysteresis in $\Delta L/L$ at 300K; this is well known [7]. The observation of even larger hysteresis at 77K is new and suggests a slowing down of the domain dynamics at this temperature. We speculate that the large hysteresis at 77K may be caused by the creation and motion of domain walls, processes that are activated and slow down at lower temperature. At 4.2K, the hysteresis has nearly vanished and the measured strain exhibits a nearly linear dependence on the piezo bias. The near absence of hysteresis at 4.2K suggests that we are in a regime where the domain walls are essentially fixed. Regardless of their origin, our data presented in Fig. 1, and in particular the presence or absence of hysteresis and linearity, should find use for the design and operation of scanning microscopes that are based on PZT actuators. Second, the ratio $[(\Delta L/L)_x / (-\Delta L/L)_y]$ is about 2 at 300 and 77K, a value consistent with the Poisson ratio expected for the deformation of a homogeneous, isotropic material under uniaxial stress when the total volume is constant. For the 4.2K data, however, this ratio is about 2.6; we do not have an explanation for this observation [8].

Next in our experiments, we thinned a 4x5mm$^2$, 0.4mm-thick GaAs wafer to about 0.1mm and glued it, using the same two-component epoxy, to the face of a similar piezo rod [9]. We then glued two strain gauges, both along the $x$ direction, one on top of the wafer and another one on the opposite face of the rod and directly on the piezo surface; we denote these by SG$_W$ and SG$_P$ (wafer and piezo strain gauges), respectively (see Fig. 2 inset). Our measurements of strain versus piezo bias based on these two gauges revealed the following. At 300K, SG$_P$ showed a similar $(\Delta L/L)_x$ as in Fig. 1(a) but SG$_W$ exhibited about a factor of 2 to 3 smaller $(\Delta L/L)_x$ and excessive creep and drift. These observations imply that at 300K the wafer did not fully follow the strain commanded by the piezo, possibly because of the weakness of the epoxy. At 77K and below, however, the strains measured by the two gauges were the same to within 2%, indicating that at low temperatures the piezo strain was fully transmitted as stress to the GaAs wafer. As an example of these results, in Fig. 2 we show $(\Delta L/L)_x$, measured by SG$_W$ for piezo biases ranging from −300V to +300V at 4.2K [10]. These data demonstrate that uniaxial stress, resulting in strains up to about $\pm 2 \times 10^{-4}$, can be applied to the semiconductor wafer *in-situ* [11]. Note also the near linearity of the applied stress with the piezo bias.

We now demonstrate how the stress provided by the piezo can be used to tune the electronic properties of a 2D electron system [1]. We used a modulation-doped sample, grown on an undoped GaAs (001) wafer via molecular beam epitaxy, in which the 2D electrons are confined to an 11nm-wide AlAs quantum well, bounded by AlGaAs barriers. The details of sample structure and its electronic properties are reported elsewhere [12]; here we give a brief summary. In bulk AlAs, electrons occupy the conduction band minima (valleys) at the X points of the Brillouin zone to form a Fermi surface that consists of six half-ellipsoids (three full ellipsoids) in the first zone. In the case of 2D electrons in an AlAs quantum well wider than ~ 5nm, because of the lattice mismatch between AlAs and GaAs and the resulting strain, the ellipsoids with their major axes lying in the plane are lower in energy and are occupied [12,13]. Denoting the growth direction as *z*, the 2D electrons in our 11nm-wide AlAs quantum well then occupy two elliptic Fermi contours with their major axes along the [100] and [010] crystallographic directions in the 2D plane, as schematically shown in Fig. 3.

In our experiments, using photolithography, we fabricated a sample in the shape of an L-shaped Hall bar with its two axes aligned with [100] and [010]. We then glued the sample to the side of the piezo rod so that [100] was along the length of the piezo rod (*x* direction). We cooled the sample in a top-loading dilution refrigerator and measured its longitudinal ($R_{xx}$ and $R_{yy}$) and Hall ($R_{xy}$ and $R_{yx}$) resistance as a function of a perpendicular magnetic field, *B*, applied in the *z* direction. Application of uniaxial stress along $x$ splits the energies of the conduction band ellipses and changes their electron occupations [see Fig. 3(b)]. Note that the total 2D electron density remains constant, as confirmed by our quantum Hall measurements. Several features of our data, such as the anisotropic sample resistance along $x$ and $y$ ($R_{xx}$ and $R_{yy}$) at zero magnetic field, provide evidence for the stress-induced splitting of the valleys. We will describe these in detail elsewhere [14]; here we confine ourselves to a simple but rather dramatic manifestation of the valley-splitting in the magneto resistance data. In Fig. 4(a) we show the change in sample resistance, $\Delta R_{xx}$, measured as a function of the applied piezo bias, at a fixed 2D density ($n = 6.07 \times 10^{11}$ cm$^{-2}$) and at the fixed $B = 1.78$T, corresponding to the Landau level filling factor $\nu = 14$. Clear oscillations of $\Delta R_{xx}$ as a function of piezo bias are seen. As we explain below, these oscillations come about because the stress-induced valley-splitting causes pairs of quantized energy levels of the 2D electron system in $B$ to cross at the Fermi level.

As schematically shown in Fig. 4(b), there are three main energies in our system. The magnetic field $B$ splits the energy into a set of Landau levels, separated by the cyclotron energy ($E_C$). Because of the electron spin, each of these levels is further split into two levels, separated the Zeeman energy ($E_z$). In the absence of any valley splitting, these energy levels in our AlAs 2D electron system should be two-fold degenerate. When we stress the sample, we remove this degeneracy and increase the valley-splitting with increasing stress, as illustrated in Fig. 4(b) [15]. This figure shows that at certain values of stress the energy levels corresponding to different valley- and spin-split Landau levels coincide at the Fermi level. At such "coincidences", the measured $R_{xx}$ minimum becomes weaker or vanishes all together [16]. Note in Fig. 4(b) that we expect the weakening/strengthening of the $R_{xx}$ minimum to be a periodic function of the applied stress, or the piezo bias, since it happens whenever the valley-splitting is an even multiple of $E_C$. In a separate report [14], we show that from similar data taken at different filling factors, we can in fact deduce values for the deformation potential of AlAs X-point conduction valleys as well as the enhancement of the valley-splitting with $B$.

In summary, we report measurements of strain versus bias for a PZT piezo in a wide temperature range. We then show how this piezo can be used to apply uniaxial stress at low temperatures in the plane of a GaAs wafer containing 2D electrons in a modulation doped AlAs sample. The results demonstrate that the stress can be continuously changed *in-situ* to dramatically modify the electronic properties of the 2D electron system.

We thank the NSF for support. M.S. also thanks the Alexander von Humboldt Foundation for support, and J. P. Kotthaus and D. Vanderbilt for helpful discussions. Part of this work was performed at the NSF-supported National High Magnetic Field Laboratory in Tallahassee, Florida; we thank E. Palm and T. Murphy for technical assistance.

**References:**

1. For stress-induced valley-splitting in Si-MOSFET 2D electrons, see G. Dorda, J. Appl. Phys. **42**, 2053 (1971), and D. C. Tsui and G. Kaminsky, Surf. Sci. **98**, 400 (1980).
2. For previous reports of displacement/bias for piezoelectric tubes in the 4 to 300K temperature range, see K. G. Vandervoort *et al.*, Rev. Sci. Instrum. **64**, 896 (1993); **65**, 3862 (1994), and D. S. Paik *et al.*, J. Mat. Sci. **34**, 469 (1999). These reports, however, do not include any hysteresis data.
3. Part No. PSt 150/5x5x7, from Piezomechanik, Munich, Germany.
4. We used two types of strain gauges: (1) "Advance" (Ni 45%, Cu 55%), part No. SG-2/350-LY11, from Omega Engineering, and (2) "Karma" (Ni 74%, Cr 20%, Al 3%, Fe 3%), part No. WK-06-062TT-350, from Vishay Micromeasurements Group. For each type, the gauge's resistance change, $\Delta R/R$, divided by its sensitivity factor, gives the strain defined as $\Delta L/L$ where $L$ is the length along the direction of the gauge axis. The Karma gauges have the advantage that corrections to their sensitivity factor, due to changes in temperature and transverse strain, are smaller and better known. Such corrections can be up to 20% (for the Advance gauges at low temperatures), and have been included in the strain data reported here.
5. Part No. 45640 "plus endfest 300", from UHU, Buehl, Germany. We cured the epoxy at 80C for 60 minutes.

6. We also have data at 0.30K and 0.03K; these are similar to the 4.2K data to within our experimental resolution.

7. See, e.g., R. C. Barrett and C. F. Quate, Rev. Sci. Instrum. **62**, 1393 (1991).

8. A similarly large ratio (~ 3) was also reported at 50K for (Ba, Sr)$TiO_3$ ceramics by D. S. Paik *et al.* (Ref. 2).

9. The use of piezo materials for applying a small AC stress to a semiconductor crystal to modulate its optical properties was reported in W. E. Engeler *et al.*, Phys. Rev. Lett. **14**, 1069 (1965).

10. Data of Figs. 1 and 2 were taken on two different piezo rods with slightly different strain versus bias characteristics.

11. Using the results of P. Lefebvre, B. Gil, H. Mathieu, and R. Planel, Phys. Rev. **B40**, 7802 (1989), we estimate that, at a piezo bias of 300V in Fig. 2, the components of the tensile stress applied to the GaAs crystal are $\sigma_x$ = 210 bar and $\sigma_y$ = 23 bar in our experiments.

12. E. P. De Poortere, Y. P. Shkolnikov, E. Tutuc, S. J. Papadakis, and M. Shayegan, Appl. Phys. Lett. **80**, 1583 (2002).

13. K. Maezawa, T. Mizutani, and S. Yamada, J. Appl. Phys. **71**, 296 (1992).

14. Y. P. Shkolnikov, K. Vakili, E. P. De Poortere, M. Shayegan, and K. Karrai, unpublished. Because of finite residual stress during sample cooldown, we needed a piezo bias of about 34V [vertical arrow in Fig. 4(a)] to attain the zero-stress condition in our experiment.

15. AlAs 2D electrons exhibit a nearly linear enhancement of valley splitting with *B* [Y. P. Shkolnikov, E. P. De Poortere, E. Tutuc, and M. Shayegan, Phys. Rev. Lett. **89**, 226805 (2002)]. This enhancement is ignored in the schematic energy level diagram shown in Fig. 4(b).

16. F. F. Fang and P. J. Stiles, Phys. Rev. **174**, 823 (1968).

**Fig. Captions:**

**Fig. 1**

Longitudinal $(\Delta L/L)_x$ and transverse $(\Delta L/L)_y$ strain vs. the bias applied to the piezo rod. Data are shown for both upsweep and downsweep of piezo bias.

**Fig. 2**

Longitudinal strain vs. piezo bias, measured at 4.2 K, with a strain gauge ($SG_W$) glued on top of a 0.1mm-thick GaAs wafer that is in turn glued to the piezo (inset). Data are shown for up and down directions of piezo bias sweeps.

**Fig. 3**

(a) Schematic illustration of occupied conduction band valleys in an 11nm-wide AlAs quantum well in the absence of uniaxial in-plane stress. (b) Application of uniaxial stress along [100] splits the energies of the valleys and therefore their occupation.

**Fig. 4**

(a) Change in resistance of an AlAs 2D electron system at filling factor $\nu=14$ ($B=1.78$T) and at $T=50$mK as a function of bias applied to the piezo rod. The vertical arrow indicates the position of zero stress. (b) Schematic energy level diagram for the AlAs 2D electrons in the presence of a fixed $B$, ignoring $B$-induced enhancement of valley splitting.

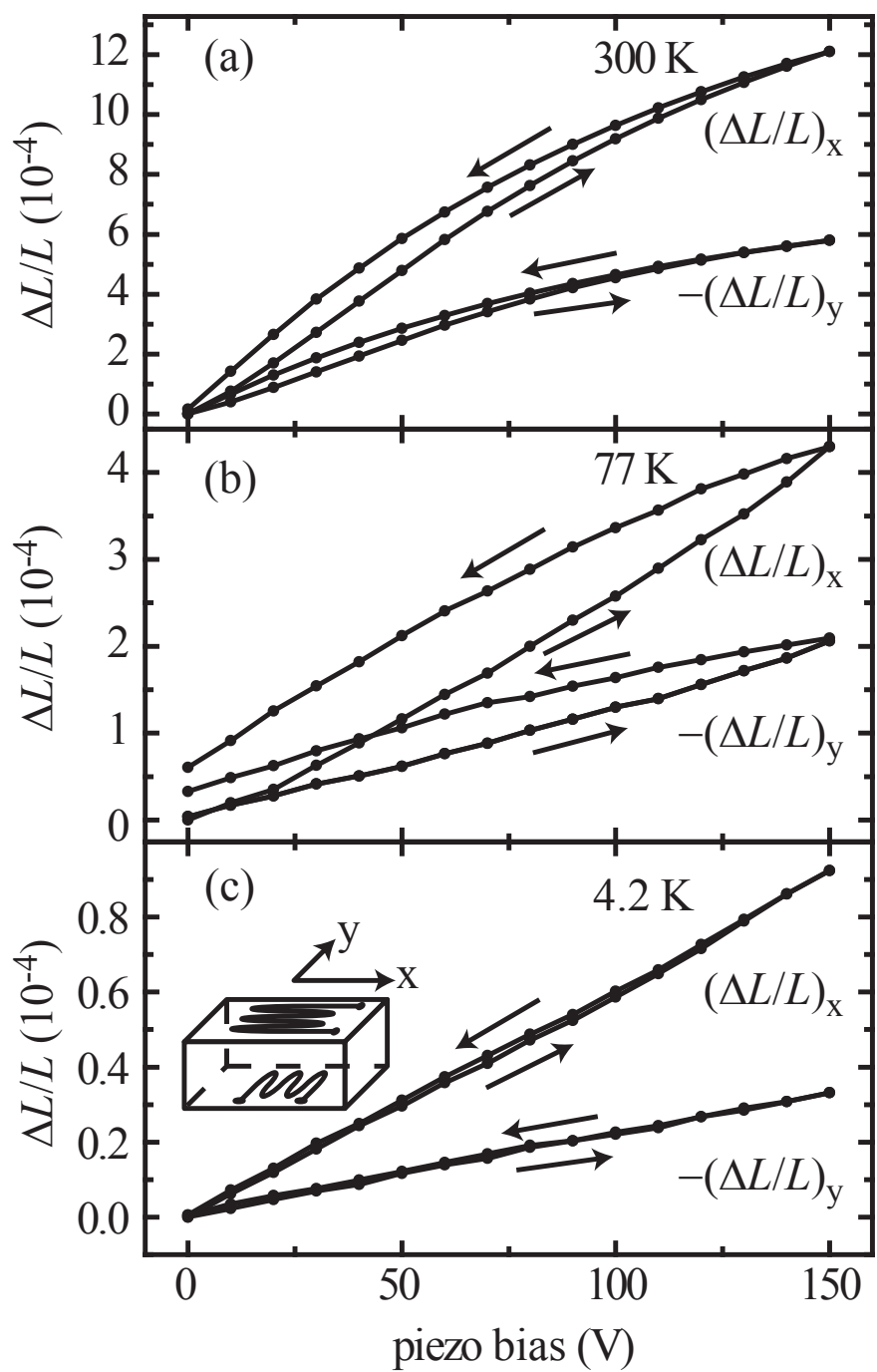

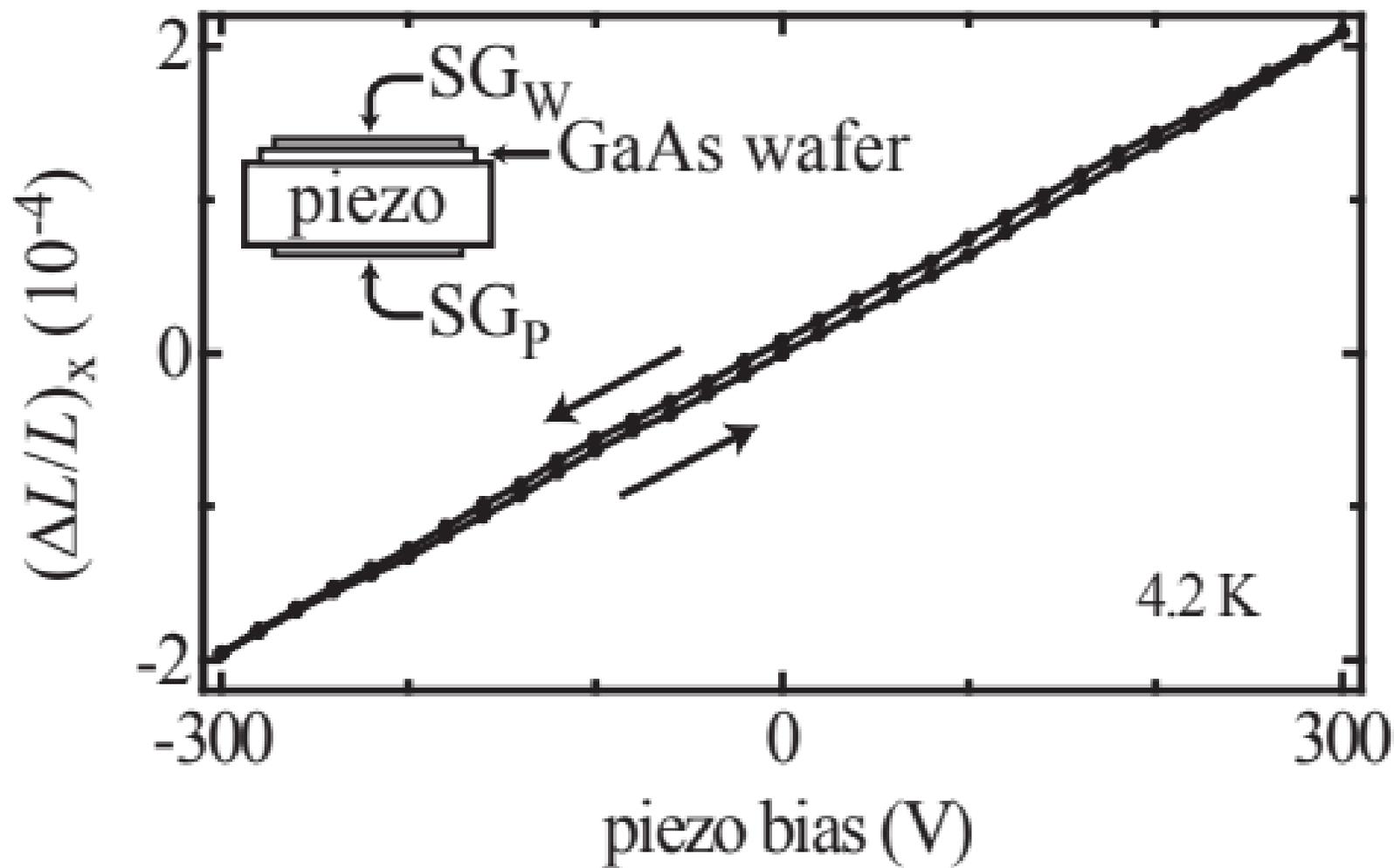

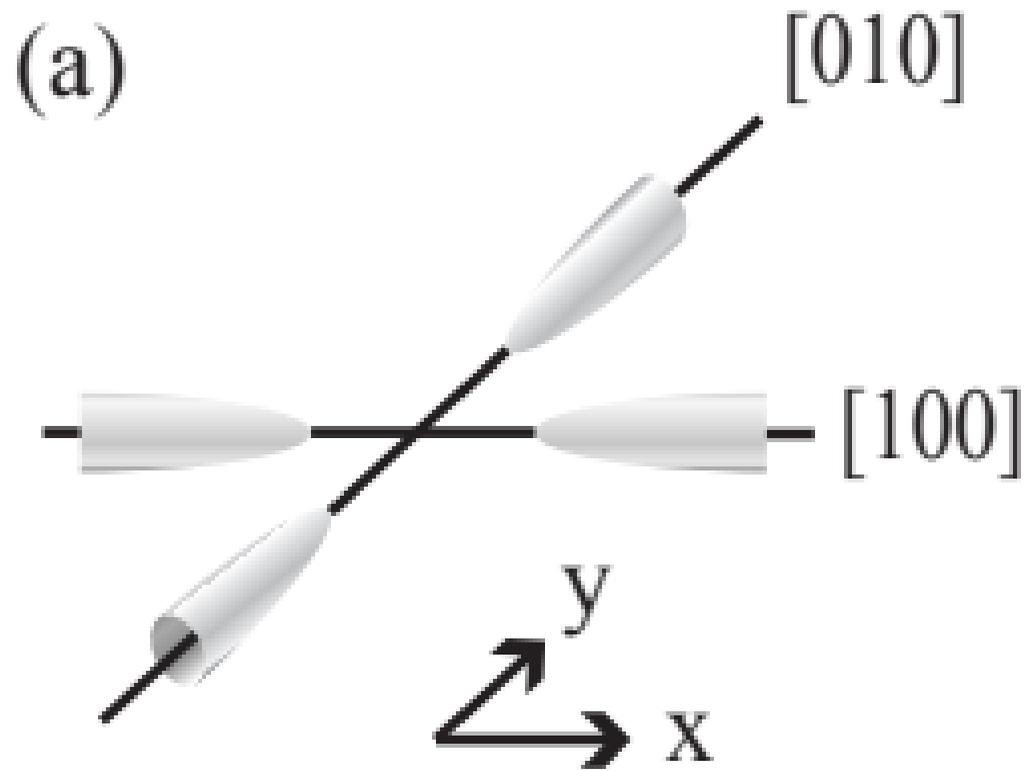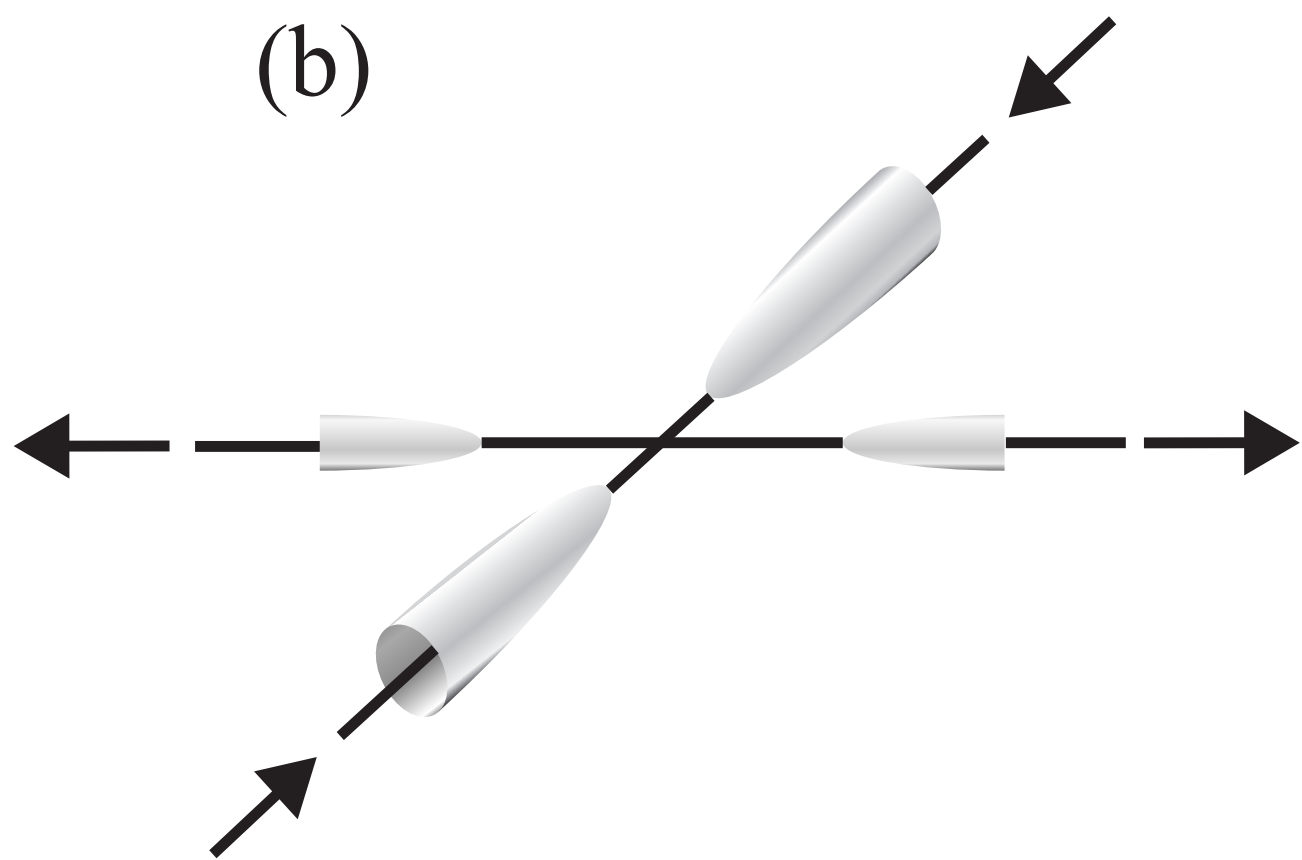

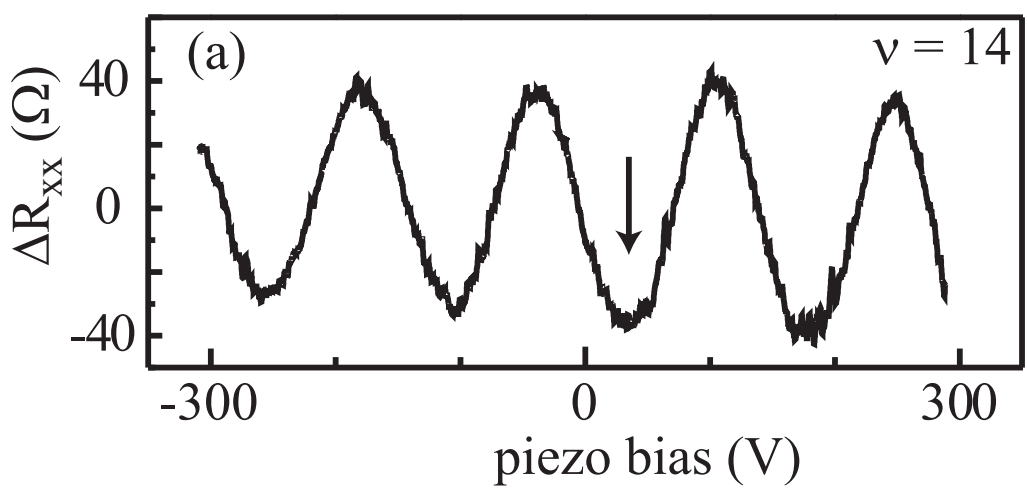

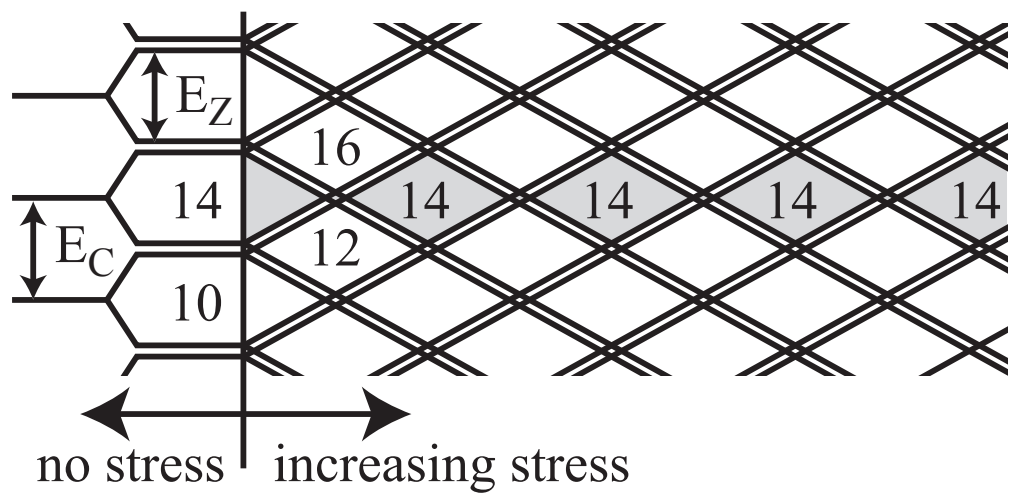